\input{aipcheck.tex}
\documentclass{aipproc}
\layoutstyle{6x9}

\begin{document}
\def\dsl{\,\raise.15ex\hbox{/}\mkern-13.5mu D} 
\def\xslash#1{{\rlap{$#1$}/}}
\def\openone{\leavevmode\hbox{\small1\kern-4.2pt\normalsize1}}
\def\ssqr#1#2{{\vbox{\hrule height #2pt
      \hbox{\vrule width #2pt height#1pt \kern#1pt\vrule width #2pt}
      \hrule height #2pt}\kern- #2pt}}
\def\sqr{\mathchoice\ssqr8{.4}\ssqr8{.4}\ssqr{5}{.3}\ssqr{4}{.3}}
\def\onedot{\makebox(0,0){$\scriptstyle 1$}}
\def\twodot{\makebox(0,0){$\scriptstyle 2$}}
\def\threedot{\makebox(0,0){$\scriptstyle 3$}}
\def\fourdot{\makebox(0,0){$\scriptstyle 4$}}
\def\Tr{{\rm Tr}}
\def\onebox{{\vbox{\hbox{$\sqr\thinspace$}}}}
\def\twobox{{\vbox{\hbox{$\sqr\sqr\thinspace$}}}}
\def\threebox{{\vbox{\hbox{$\sqr\sqr\sqr\thinspace$}}}}
\def\nbox{\hbox{$\sqr\sqr\sqr\sqr\raise2.7pt\hbox{$\,\cdot\cdot\cdot
\cdot\cdot\,$}\sqr\sqr\sqr\thinspace$}}
\def\nboxE{\vbox{\hbox{$\sqr\sqr\sqr\raise2.7pt\hbox{$\,\cdot\cdot\cdot
\cdot\cdot\,$}\sqr\sqr\sqr\sqr$}\nointerlineskip 
\kern-.2pt\hbox{$\sqr\sqr\sqr\raise2.7pt\hbox{$\,\cdot\cdot\cdot
\cdot\cdot\,$}\sqr$}}}
\def\nboxF{\vbox{\hbox{$\sqr\sqr\sqr\sqr\raise2.7pt\hbox{$\,\cdot\cdot\cdot
\cdot\cdot\,$}\sqr\sqr$}\nointerlineskip 
\kern-.2pt\hbox{$\sqr\sqr\sqr\sqr\raise2.7pt\hbox{$\,\cdot\cdot\cdot
\cdot\cdot\,$}\sqr$}}}

\title
[Ferromagnets and antiferromagnets in the effective Lagrangian perspective]
{Ferromagnets and antiferromagnets in the effective Lagrangian perspective}

\author
{Christoph P. Hofmann}{
  address={Instituto de F{\'\i}sica, Universidad Aut\'onoma de San Luis
  Potos{\'\i}, Alvaro Obreg\'on 64, San Luis Potos{\'\i}, S.L.P.\ 78000,
  Mexico},}

\copyrightyear  {2001}

\begin{abstract}
Nonrelativistic systems exhibiting collective magnetic behavior are analyzed within the
framework of effective Lagrangians. The method, which formulates the dynamics of the
system in terms of Goldstone bosons, allows to investigate the consequences of spontaneous
symmetry breaking from a unified point of view. Analogies and differences with respect to
the Lorentz-invariant situation (chiral perturbation theory) are pointed out. We then
consider the low-temperature expansion of the partition function both for ferro- and
antiferromagnets, where the spin waves or magnons represent the Goldstone bosons of the
spontaneously broken symmetry $O(3) \to O(2)$. In particular, the low-temperature series
of the staggered magnetization for antiferromagnets and the spontaneous magnetization for
ferromagnets are compared with the condensed matter literature.
\end{abstract}

\date{\today}

\maketitle
In condensed matter physics, spontaneous symmetry breaking is a common phenomenon and
effective field theory methods are widely used in this domain. Only recently, however,
has chiral perturbation theory -- the effective theory of the strong interactions -- been
extended to such nonrelativistic systems.\cite{Leutwyler NRD} The method applies to any
system where the Goldstone bosons are the only excitations without energy gap. The
essential point is that the properties of these degrees of freedom and their mutual
interactions are strongly constrained by the symmetry inherent in the underlying model --
the specific nature of the underlying model itself, however, is not important.

In the following presentation, our interest is devoted to the low-energy
analysis of nonrelativistic systems, which exhibit collective magnetic
behavior. The Heisenberg Hamiltonian is invariant under a simultaneous
rotation of the spin variables, described by the symmetry group G = O(3),
whereas the ground states of ferro- and antiferromagnets break this symmetry
spontaneously down to H = O(2). The corresponding Goldstone modes are referred to
as spin waves or magnons. Note that, in contrast to the relativistic version of
Goldstone's theorem, the theorem does now neither specify the exact form of the
dispersion relation at large wavelengths, nor does it determine the number of different
Goldstone {\it particles}: these features of the Goldstone degrees of freedom are not
fixed by symmetry considerations alone -- rather, in the case of a Lorentz-noninvariant
ground state, they depend on the specific properties of the corresponding
nonrelativistic systems. Only the number of real Goldstone {\it fields} turns
out to be universal, given by the dimension of G/H.

Indeed, it is well known that the structure of the ferromagnetic dispersion relation is
quite different from the antiferromagnetic one: at large wavelengths, the
former takes a {\it quadratic} form, whereas the latter follows a {\it linear}
law. The mechanism which leads to this pattern and, at the same time, explains
the different number of independent magnon states -- {\it one} for a
ferromagnet, {\it two} for an antiferromagnet -- is understood. Remarkably, in the
framework of the effective description, the difference is related to the value of a
single observable, the spontaneous magnetization $\Sigma$.\cite{Leutwyler NRD, Hofmann spin
wave}

In the leading order effective Lagrangian of a {\it ferromagnet}, the spontaneous
magnetization shows up as a coupling constant associated with a topological term involving
a single time derivative, 
\begin{equation}
\label{LeffFerro}
{\cal L}_{eff}^{F} \ = \ \Sigma \, \frac{{\partial}_0 U^1U^2 - {\partial}_0
U^2U^1}{1 + U^3} \ + \ {\Sigma}f^i_0 U^i \ - \ \mbox{$ \frac{1}{2}$} F^2 D_rU^iD_rU^i .
\end{equation}
In the above notation, the two real components of the magnon field, $U^a(a=1,2)$ have been
collected in a three-dimensional unit vector $U^i = (U^a, U^3)$. The quantity $f^i_0$
involves the magnetic field $H$: $f^i_0 = \mu H \delta^i_3$. At leading order of the
low-energy expansion, the ferromagnet is thus characterized by two low-energy coupling
constants, $\Sigma$ and $F$. Note that the corresponding equation of motion (Landau-Lifshitz
equation) is of the Schr\"odinger type: first order in time, but second order in space. As
only positive frequencies occur in its Fourier decomposition, a complex field is required to
describe one particle -- in a ferromagnet there exists only {\it one} type of spin-wave
excitation exhibiting a quadratic dispersion law.

The ground state of an antiferromagnet, on the other hand, does not exhibit
spontaneous magnetization, such that the leading order effective Lagrangian takes the form
\begin{equation}
\label{LeffAF}
{\cal L}_{eff}^{AF} \, = \, \mbox{$ \frac{1}{2}$} F^2_1
D_0U^iD_0U^i - \mbox{$ \frac{1}{2}$} F^2_2 D_rU^iD_rU^i + {\Sigma}_s \mu h^i U^i \, , \quad
D_{\mu}U^i = {\partial}_{\mu}U^i + \varepsilon_{ijk} f^j_{\mu}U^k .
\end{equation}
Note that the anisotropy field $\vec h$ couples to the staggered magnetization ${\Sigma}_s$.
The corresponding equation of motion now is of second order both in space and in time, its
relativistic structure determining the number of independent magnon states: the Fourier
decomposition contains both positive and negative frequencies, such that a single real
field suffices to describe one particle. Accordingly, there exist {\it two} different
types of spin-wave excitations in an antiferromagnet -- as is the case in
Lorentz-invariant theories, Goldstone fields and Goldstone particles are in one-to-one
correspondence. These low-energy excitations follow a linear dispersion relation, with the
velocity of light replaced by the spin-wave velocity $v= F_2 / F_1$. As is commonly done with
the velocity of light in relativistic theories, we may put the spin-wave velocity to one. In
this "$\hbar = v = 1$"-system the two coupling constants $F_1$ and $F_2$ then coincide:
$F_1 = F_2 \equiv {\cal F}$.

Accordingly, the low-energy properties of ferromagnets and antiferromagnets are quite
different. As an illustration, let us consider the low-temperature expansion for the
corresponding order parameters, the spontaneous and staggered magnetization, respectively.

The order parameter for an O(N) antiferromagnet, the staggered magnetization, is given by the
derivative of the free energy density with respect to the anisotropy field,\cite{Hofmann AF}
${\Sigma}_s (T) \, = \, -  \, \partial z / \partial h$:
\begin{eqnarray}
{\Sigma}_s(T) \ = \ {\Sigma}_s \; \Bigg\{1 \; - \; \frac{N\!-\!1}{24}
\frac{T^2}{{\cal F}^2} \; 
- \; \frac{(N\!-\!1)\,(N\!-\!3)}{1152} \, \frac{T^4}{{\cal F}^4} \hspace{3cm}\\
\hspace{3cm} - \; \frac{(N\!-\!1)\,(N\!-\!2)}{1728} \, \frac{T^6}{{\cal F}^6} \,
\ln{\frac{T_{\Sigma}}{T}} \, + {\cal O} (T^8) \Bigg\} . \nonumber
\end{eqnarray}
The terms of order $T^0, T^2, T^4$ and $T^6$ arise from tree-, one-loop, two-loop and
three-loop graphs, respectively. Up to and including $T^6$, the coefficients are determined
by the constant $\cal F$ which thus sets the scale of the expansion. The logarithm only
shows up at order $T^6$: the scale $T_{\Sigma}$ involves next-to-leading order coupling
constants.

Let us first consider the particular case N=4. The two groups O(4) and O(3) are locally
isomorphic to SU(2) $\times$ SU(2) and SU(2), respectively. Hence, the above three-loop
formula referring to the order parameter of an O(4) antiferromagnet in zero external field
in fact describes the low-temperature expansion of the quark condensate of massless QCD with
two flavors. This nicely illustrates the concept of {\it universality}: in the construction
of effective Lagrangians only the mathematical structure of the groups G and H, associated
with the spontaneously broken symmetry, is relevant, whereas the specific properties of the
underlying model merely manifest themselves in the numerical values of the coupling
constants.

Remarkably, for N=3, the $T^4$-term in the above formula drops out, such that we end up with
the following low-temperature series for the staggered magnetization of the O(3)
antiferromagnet:
\begin{eqnarray}
\label{OrdPar3}
{\Sigma}_s(T) \ = \ {\Sigma}_s \; \Bigg\{1 \; - \; \frac{1}{12}
\frac{(k_{B} T)^2}{ \hbar v  {\cal F}^2} \; - \; \frac{1}{864}
\frac{(k_{B} T)^6}{{\hbar}^3 v^3 {\cal F}^6} \,
\ln{\frac{T_{\Sigma}}{T}} \, + {\cal O} (T^8) \Bigg\} \, .
\end{eqnarray}
Note that we have restored the dimensions: $k_B$ is Boltzmann's constant and $v$ is the
spin-wave velocity.

The microscopic calculation agrees with the above effective expansion up to order $T^2$,
provided that the two coupling constants $\cal F$ and ${\Sigma}_s$ are identified as
\begin{equation}
\label{EffConstMic}
{\cal F}^2 \; = \; \frac{S - \sigma}{\sqrt{2z}} \, \frac{\hbar v}{a^2}
\; = \; 2S (S - \sigma) \, \frac{|J|}{a} \, , \qquad
{\Sigma}_s \; = \; \frac{g {\mu}_{B}(S - \sigma)}{a^3} \, .
\end{equation}
The expression involves the following quantities: the exchange integral ($J$), the highest
eigenvalue of the spin operator $S^3_n$ ($S$), the number of nearest neighbors of a given
lattice site ($z$), the length of the unit cell ($a$),
the Land\'e factor ($g$), the "Anderson factor" ($\sigma$) and the Bohr magneton ($\mu_B$).
Note that the spin-wave velocity $v$ is given by the following combination of microscopic
quantities,
\begin{equation}
\label{spinwaveVelocity}
v = \, 2|J|S\sqrt{2z} \, a / \hbar \, .
\end{equation}
The scale of the low-temperature expansion is set by ${\cal F} \sqrt{\hbar v}$ -- let
us briefly estimate its value. Written in terms of the exchange integral $J$,
we obtain
\begin{equation}
{\cal F} \sqrt{\hbar v} \, = \, 2|J| S \sqrt{(S - \sigma) \sqrt{2z}} \, .
\end{equation}
Now, for a simple cubic lattice ($z \! = \! 6, \sigma \! = \! 0.078$) and for
$S = 1/2$, the double square root on the right hand side is approximately equal
to one, such that we end up with ${\cal F} \sqrt{\hbar v} \approx |J|$. Typically,
the exchange integral for antiferromagnets is around $|J| \approx 10^{-3} eV$,
and the scale ${\cal F} \sqrt{\hbar v}$ thus of the same order of magnitude. This
is to be contrasted with the situation in QCD, where the relevant quantity,
$F_{\chi} \sqrt{\hbar c}$, takes the value $92 MeV$ -- the respective scales in the
two theories thus differ in about eleven orders of magnitude.

As far as subleading terms in the expansion of the staggered magnetization are
concerned, it is well known that a $T^4$-contribution is absent: the spin-wave
interaction only manifests itself at higher orders. However, the logarithmic dependence
on the temperature is not found in a microscopic calculation. We conclude that it is
extremely difficult to calculate the corrections of order $T^6$ in the framework of a
microscopic theory.

Let us now turn to the ferromagnet. The low-temperature expansion for the spontaneous
magnetization is given by the derivative of the free energy density with respect to the
magnetic field and takes the
form\cite{Hofmann Ferro}
\begin{equation}
\Sigma(T) \, / \, \Sigma \ = \ 1 \, - \, {\alpha}_0 T^{\frac{3}{2}} \, - \, {\alpha}_1
T^{\frac{5}{2}} \, - \, {\alpha}_2 T^{\frac{7}{2}} \, - \, {\alpha}_3 T^4 \,
+ \, {\cal O} (T^{{\frac{9}{2}}}) .
\end{equation}

The coefficients $\alpha_n$ are independent of the temperature and involve the various
coupling constants occurring in the effective Lagrangian, which phenomenologically
parametrize the microscopic detail of the system.

In the above series, half-integer powers of the temperature correspond to {\it
noninteracting} magnons; these contributions can be absorbed into a redefinition of the
dispersion relation. Remarkably, the leading term describing the magnon-magnon {\it
interaction} (two-loop graph) is of order $T^4$, and clearly confirms
Dyson's microscopic calculation.\cite{Dyson} The effective Lagrangian technique, however,
proves to be more efficient than conventional condensed matter methods as the analysis can
be carried to higher orders: the calculation shows that the next interaction term, which
arises at the three-loop level, is of order $T^{9/2}$.

The effective Lagrangian method is also more transparent, since it addresses the problem
from a model-independent point of view based on symmetry -- at large wavelengths, the
microscopic structure of the system only manifests itself in the numerical values of a few
coupling constants.


\begin{thebibliography}{100}

\bibitem{Leutwyler NRD}
H. Leutwyler, Phys. Rev. D {\bf 49}, 3033-3043 (1994).

\bibitem{Hofmann spin wave}
C.P. Hofmann, Phys. Rev. B {\bf 60}, 388-405 (1999).

\bibitem{Hofmann AF}
C.P. Hofmann, Phys. Rev. B {\bf 60}, 406-413 (1999).

\bibitem{Hofmann Ferro}
C.P. Hofmann, cond-mat/0106492, to appear in Phys. Rev. B.

\bibitem{Dyson}
F.J. Dyson, Phys. Rev. {\bf 102}, 1217-1230, 1230-1244 (1956).

\end{thebibliography}
\end{document}